\documentclass[11pt,showpacs,preprintnumbers,superscriptaddress,amsmath,amssymb,nofootinbib]{revtex4}

\usepackage{graphicx}
\usepackage{dcolumn}
\usepackage{bm}
\usepackage{amssymb}
\usepackage{amsmath}
\usepackage{epsfig}    
\usepackage{color}
\usepackage{slashed}
\usepackage{float}
\newcolumntype{I}{!{\vrule width 1.3pt}}
\begin{document} 
\title{Phenomenology of the Georgi-Machacek model\\ at future electron-positron colliders}

\author{Cheng-Wei Chiang}
\email{chengwei@ncu.edu.tw}
\affiliation{Department of Physics and Center for Mathematics and Theoretical Physics, 
National Central University, Chungli, Taiwan 32001, ROC}
\affiliation{Institute of Physics, Academia Sinica, Taipei, Taiwan 11529, ROC}
\affiliation{Physics Division, National Center for Theoretical Sciences, Hsinchu, Taiwan 30013, ROC}
\author{Shinya Kanemura}
\email{kanemu@sci.u-toyama.ac.jp}
\affiliation{Department of Physics, University of Toyama, 3190 Gofuku, Toyama 930-8555, Japan}
\author{Kei Yagyu}
\email{K.Yagyu@soton.ac.uk}
\affiliation{School of Physics and Astronomy, University of Southampton, Southampton, SO17 1BJ, United Kingdom}


\begin{abstract}
We study the phenomenology of the exotic Higgs bosons in the Georgi-Machacek model at future electron-positron colliders such as the 
International Linear Collider (ILC), assuming the collision energies of 500~GeV and 1~TeV.  
We show that the existence of the neutral and singly-charged Higgs bosons 
in the 5-plet representation under the custodial $SU(2)_V$ symmetry
can be readily identified by studying various energy and invariant mass distributions of the $W^+W^-Z$ final state.  Moreover, their masses can be determined with sufficiently high precision to test the mass degeneracy, a feature due to the custodial symmetry of the model.  A synergy between such searches at the ILC and the doubly-charged Higgs search at the LHC will make the 5-plet Higgs boson study more comprehensive.
\end{abstract}
\date{\today}
\maketitle

\section{introduction }

Since the discovery of the 125-GeV Higgs boson at the CERN Large Hadron Collider (LHC), efforts have been made to study in more detail its properties, particularly how it interacts with other particles in the standard model (SM).  
Such an endeavor has partly the mission of exploring whether there is an extended Higgs sector and, if so, how it will help us understand 
the Nature.  
If certain new physics models, such exotic Higgs bosons may hold answers to some long-lasting questions in particle physics, such as the origin of neutrino mass, the identity of dark matter, and the realization of a strong first-order phase transition for electroweak baryogenesis.

As an extension of the SM in the Higgs sector, the Georgi-Machacek (GM) model~\cite{Georgi:1985nv,Chanowitz:1985ug} has some unique and desirable properties in comparison with other Higgs-extended models.  The GM model has a doublet field $\phi$ with hypercharge $Y = 1/2$ and a triplet field composed of a complex triplet $\chi$ with $Y = 1$ and a real triplet $\xi$ with $Y = 0$.  By starting with a Higgs potential with the custodial $SU(2)_V$ 
symmetry and the vacuum alignment between the complex and real triplets, the model preserves the electroweak rho parameter $\rho = 1$ at tree level while allowing the possibility of a triplet vacuum expectation value (VEV), $v_\Delta$, as large as up to a few tens of GeV.  Through appropriate Yukawa couplings with the leptons, such a VEV can contribute to the mass of neutrinos {\it {\`a} la} type-II seesaw mechanism~\cite{typeII}.  The couplings between the triplet field and leptons lead to lepton number-violating processes and possibly even lepton flavour-violating ones.  Another consequence of the custodial symmetry is mass degeneracy within each Higgs multiplet~\cite{Gunion:1989ci,Chiang_Yagyu_GM}.  It has been demonstrated to be possible to determine the mass and check this property at the Large Hadron Collider (LHC)~\cite{Chiang_Yagyu_GM}.

A distinctive feature of the model is that, among the exotic Higgs bosons, there is a doubly-charged Higgs boson, $H_5^{\pm\pm}$, in the 5-plet representation.  Such a particle can lead to phenomenologically prominent and interesting signatures at colliders: 
decays into a pair of like-sign leptons or $W$ bosons, depending on the magnitude of $v_\Delta$.  Due to the mixing between the Higgs doublet and triplet fields, 
the couplings between the SM-like Higgs boson $h$ and the weak gauge bosons can be stronger than their 
SM values~\cite{Logan:2010en,Falkowski:2012vh,Chang:2012gn,Chiang:2013rua,hvv1} 
and lead to discriminative phenomena~\cite{Chiang:2013rua,Godfrey:2010qb,Chiang:2012dk,Englert:2013zpa,fingerprint}.  
Another consequence of significant mixing between the Higgs doublet and triplet fields is the possibility of a strong first-order phase transition for electroweak baryogenesis in some parameter space~\cite{Chiang:2014hia}.  
Besides, the model has a tree-level $H_5^{\pm}W^{\mp}Z$ coupling, which is known to be small in multi-doublet models because they appear only at loop levels~\cite{hwz1,hwz2,hwz3}.

As mentioned above, the doubly-charged Higgs boson can be searched using the like-sign dilepton and diboson modes.  LHC results on the former in the past few years had placed a lower bound of about 400~GeV on its mass for some generic benchmark points of the model~\cite{Chatrchyan:2012ya,ATLAS:2012hi,Aad:2012xsa}.  The like-sign diboson production with leptonic decays of the $W$ bosons had also been searched for by the ATLAS Collaboration using the $4.7$ fb$^{-1}$ data at the 7-TeV run~\cite{ATLAS:2012mn}, from which a mass lower bound of $\lesssim 70$~GeV was derived~\cite{Kanemura:2014goa}.  
More recently, the ATLAS also measured the total production cross section of like-sign $W$ bosons and a pair of jets in the 8-TeV run~\cite{ATLAS:2014rwa}.  
The reported production cross section in the vector boson scattering fiducial region had been employed to constrain the triplet VEV as a function of the doubly-charged Higgs mass~\cite{Chiang:2014bia}.  
Of particular interest to the current work is that the 95\%~CL upper bound on $v_\Delta$ from the LHC data~\cite{Chiang:2014bia}
is at the level of several tens of GeV, 
opening up the possibility of studying the GM Higgs bosons at lepton colliders.

Although in the case of a large triplet VEV the exotic Higgs bosons have diminishing Yukawa couplings with charged leptons, the 5-plet Higgs bosons can still be produced via productions in association with weak gauge bosons that serve as promising detection channels at lepton colliders, such as the International Linear Collider (ILC)~\cite{AguilarSaavedra:2001rg,Abe:2001wn,Abe:2001gc,Moortgat-Picka:2015yla}, the Compact Linear Collider (CLIC)~\cite{Accomando:2004sz}, and the circular electron-position collider (CEPC)~\cite{CEPC-SPPCStudyGroup:2015csa} or the electron-positron branch of the Future Circular Collider (FCC-ee).  There are a few earlier studies in this direction.  Refs.~\cite{9411324,9411333} examined the possibility of probing this sector using the uniquely featured tree-level vertex of $H_5^\pm W^\mp Z$ at high-energy $e^+e^-$ colliders.

In this paper, we concentrate on the study of how one can test the GM model at the ILC with proposed colliding energies of 500~GeV and 1~TeV.  We show that with a cleaner collider environment, it is easier to determine the 5-plet mass with high precision at the ILC than the LHC.  We point out that the vector boson associated production processes with the $W^+W^-Z$ channel can be used to determine the masses of $H_5^\pm$ and $H_5^0$.  Besides, this channel has a wider probing range in the 5-plet mass than the other related multiple weak boson modes.

The structure of this paper is as follows.  Section~\ref{sec:model} reviews the Georgi-Machacek model and, in particular, give the relations between model parameters and physical observables.  In Section~\ref{sec:5plet}, we discuss possible decay patterns of the 5-plet Higgs bosons in the model.  Branching ratios of different charged states are explicitly worked out as a function of the mass difference between the 5-plet and 3-plet Higgs bosons.  In Section~\ref{sec:production}, we give numerical results regarding the production of the 5-plet Higgs bosons at the ILC, assuming a center-of-mass energy of 500~GeV and 1~TeV.  After analyzing the SM backgrounds of several possible gauge boson final states of the exotic Higgs production in the model, we concentrate in Section~\ref{sec:reconstruction} on the $W^+W^-Z$ events and show that the energy and invariant mass distributions of a subset of these final-state particles can be used to determine the 5-plet mass.  Section~\ref{sec:discussions} discusses how the searches for the 5-plet Higgs bosons at the ILC complements the corresponding searches at the LHC.  Our summary is given in Section~\ref{sec:summary}.

\section{The model \label{sec:model}}

The Higgs sector of the GM model is composed of an isospin doublet field $\phi$ with the hypercharge\footnote{In our paper, the relation between the electric charge $Q$ and the third component isospin $T_3$ is given by $Q=T_3+Y$. } $Y=1/2$, a complex triplet field $\chi$ with $Y=1$, and a real triplet field $\xi$ with $Y=0$. 
The doublet and triplet fields can respectively be expressed in the following $SU(2)_L\times SU(2)_R$-covariant doublet and triplet forms:
\begin{align}
\Phi=\left(
\begin{array}{cc}
\phi^{0*} & \phi^+ \\
-\phi^- & \phi^0
\end{array}\right),\quad 
\Delta=\left(
\begin{array}{ccc}
\chi^{0*} & \xi^+ & \chi^{++} \\
-\chi^- & \xi^0 & \chi^{+} \\
\chi^{--} & -\xi^- & \chi^{0} 
\end{array}\right), 
\label{eq:Higgs_matrices}
\end{align}
where we use the convention that $\chi^{--}=(\chi^{++})^*$, $\chi^{-}=(\chi^{+})^*$, $\xi^-= (\xi^+)^*$ and $\phi^-= (\phi^+)^*$. 
The neutral components in Eq.~(\ref{eq:Higgs_matrices}) can be parameterized as 
\begin{align}
\phi^0&=\frac{1}{\sqrt{2}}(\phi_r+v_\phi+i\phi_i), \quad 
\chi^0=\frac{1}{\sqrt{2}}(\chi_r+i\chi_i)+v_\chi,\quad \xi^0=\xi_r+v_\xi, \label{eq:neutral}
\end{align}
where $v_\phi$, $v_\chi$ and $v_\xi$ are the VEV's for $\phi$, $\chi$ and $\xi$, respectively.

The most general Higgs potential invariant under the $SU(2)_L\times SU(2)_R\times U(1)_Y$ gauge group is given 
in terms of the matrix representations defined in Eq.~(\ref{eq:Higgs_matrices}) by 
\begin{align}
V_H&=m_1^2\text{tr}(\Phi^\dagger\Phi)+m_2^2\text{tr}(\Delta^\dagger\Delta)
+\lambda_1[\text{tr}(\Phi^\dagger\Phi)]^2+\lambda_2[\text{tr}(\Delta^\dagger\Delta)]^2
+\lambda_3\text{tr}[(\Delta^\dagger\Delta)^2]\notag\\
&+\lambda_4\text{tr}(\Phi^\dagger\Phi)\text{tr}(\Delta^\dagger\Delta)
+\lambda_5\text{tr}\left(\Phi^\dagger\frac{\tau^a}{2}\Phi\frac{\tau^b}{2}\right)
\text{tr}(\Delta^\dagger t^a\Delta t^b)\notag\\
&+\mu_1\text{tr}\left(\Phi^\dagger \frac{\tau^a}{2}\Phi\frac{\tau^b}{2}\right)(P^\dagger \Delta P)^{ab}
+\mu_2\text{tr}\left(\Delta^\dagger t^a\Delta t^b\right)(P^\dagger \Delta P)^{ab}, \label{eq:pot}
\end{align}
where $\tau^a$ and $t^a$ ($a=1,2,3$) are the $2\times 2$ (the Pauli matrices) and $3\times 3$ matrix representations of the $SU(2)$ generators, respectively.  
The matrix $P$ diagonalizes one of the adjoint representation matrices of the $SU(2)$ generator, and is explicitly expressed as
\begin{align}
P=
\begin{pmatrix}
-1/\sqrt{2} & i/\sqrt{2} & 0 \\
0 & 0 & 1 \\
1/\sqrt{2} & i/\sqrt{2} & 0
\end{pmatrix}. 
\end{align}
The soft-breaking terms with $\mu_1$ and $\mu_2$ of the $Z_2$ symmetry (under the transformations of $\Phi \to +\Phi$ and $\Delta \to -\Delta$)
in the Higgs potential are necessary to obtain the decoupling limit to the SM when
taking them into infinity. 
We note that no CP-violating term is allowed in the above potential. 
When we take $ v_\Delta^{}  \equiv v_\chi=v_\xi $, 
the $SU(2)_L\times SU(2)_R$ symmetry is reduced to the custodial $SU(2)_V$ symmetry. 
In that case, the masses of the weak gauge bosons are given by the same forms as those in the SM:
\begin{align}
m_W^2 = \frac{g^2v^2}{4},\quad m_Z^2=\frac{g^2v^2}{4\cos^2\theta_W}, 
\end{align}
where $v^2\equiv v_\phi^2+8v_\Delta^2=1/(\sqrt{2}G_F)$. 
Thus, the electroweak rho parameter $\rho \equiv m_W^2/(m_Z^2\cos^2\theta_W)$ is unity at tree level.

The component scalar fields can be classified into the irreducible representations of 5-plet, 3-plet and singlet under $SU(2)_V$. 
That is, the scalar fields from doublet $\Phi$ can be decomposed as ${\bf 2}\otimes {\bf 2} \to {\bf 3}\oplus{\bf 1}$, and 
those from the triplet $\Delta$ can be done as ${\bf 3}\otimes {\bf 3} \to {\bf 5}\oplus{\bf 3}\oplus{\bf 1}$. 
Among these $SU(2)_V$ multiplets, the 5-plet states directly become physical Higgs bosons, {\it i.e.}, $H_5=(H_5^{\pm\pm},H_5^\pm,H_5^0$).  For the two 3-plets, one of the linear combinations corresponds to physical Higgs field, {\it i.e.}, $H_3=(H_3^\pm,H_3^0$), and the other becomes the Nambu-Goldstone (NG) bosons $G^\pm$ and $G^0$ which are absorbed into the longitudinal components of the $W^\pm$ and $Z$ bosons, respectively.  Furthermore, we have two $SU(2)_V$ singlets which are mixed with each other in general, with one of the two mass eigenstates being identified as the discovered 125-GeV Higgs boson.  Because of the $SU(2)_V$ invariance, different charged Higgs boson states belonging to the same $SU(2)_V$ multiplet are degenerate in mass.

The scalar bosons in the mass eigenbasis are related to their weak eigenstates defined in Eqs.~(\ref{eq:Higgs_matrices}) and (\ref{eq:neutral}) via the following orthogonal transformations
\begin{align}
\left(
\begin{array}{c}
\phi_i\\
\chi_i
\end{array}\right)
&=U_{\text{CP-odd}}
\left(
\begin{array}{c}
G^0\\
H_3^0
\end{array}\right),~
\left(
\begin{array}{c}
\phi^\pm\\
\xi^\pm\\
\chi^\pm
\end{array}\right)
=U_\pm
\left(
\begin{array}{c}
G^\pm\\
H_3^\pm\\
H_5^\pm
\end{array}\right),~
\left(
\begin{array}{c}
\phi_r\\
\xi_r\\
\chi_r
\end{array}\right)=U_{\text{CP-even}}
\left(
\begin{array}{c}
h\\
H_1^0\\
H_5^0
\end{array}\right). 
\end{align}
The above three transformation matrices are given by 
\begin{align}
U_{\text{CP-odd}}&=
\left(
\begin{array}{cc}
c_H & -s_H \\
s_H & c_H
\end{array}\right),~
U_\pm =\left(
\begin{array}{ccc}
1 & 0 & 0\\
0 & \frac{1}{\sqrt{2}} & -\frac{1}{\sqrt{2}} \\
0&\frac{1}{\sqrt{2}} &\frac{1}{\sqrt{2}} 
\end{array}
\right)\left(
\begin{array}{ccc}
c_H & -s_H & 0\\
s_H & c_H & 0\\
0&0&1
\end{array}
\right),\notag\\
U_{\text{CP-even}}&=
\left(
\begin{array}{ccc}
1 & 0 &0\\
0 & \frac{1}{\sqrt{3}} & -\sqrt{\frac{2}{3}}\\
0 & \sqrt{\frac{2}{3}} & \frac{1}{\sqrt{3}}
\end{array}\right)
\left(
\begin{array}{ccc}
 c_\alpha & s_\alpha &0\\
 -s_\alpha & c_\alpha &0\\
0 & 0 & 1 
\end{array}\right), 
\end{align}
where $c_H=\cos\theta_H$ and $s_H=\sin\theta_H$ with $\tan\theta_H=2\sqrt{2}v_\Delta/v_\phi$. 
We also introduced $c_\alpha = \cos\alpha$ and $s_\alpha = \sin\alpha$.

The squared masses of the 5-plet Higgs bosons ($m_{H_5}^2$), the 3-plet Higgs bosons ($m_{H_3}^2$) and 
the two singlet Higgs bosons $h$ ($m_h^2$) and $H$ ($m_{H_1}^2$) are given by 
\begin{align}
m_{H_5}^2 &= \left(s_H^2\lambda_3 -\frac{3}{2}c_H^2\lambda_5\right)v^2+c_H^2M_1^2+M_2^2, \\
m_{H_3}^2 &= -\frac{1}{2}\lambda_5v^2+M_1^2, \\
m_h^2    &=(M^2)_{11}c_\alpha^2+(M^2)_{22}s_\alpha^2 -2(M^2)_{12}s_\alpha c_\alpha, \\
m_{H_1}^2  &=(M^2)_{11}s_\alpha^2+(M^2)_{22}c_\alpha^2 +2(M^2)_{12}s_\alpha c_\alpha, 
\end{align}
and the mixing angle $\alpha$ is given by
\begin{align}
\tan2\alpha =\frac{2(M^2)_{12}}{(M^2)_{22}-(M^2)_{11}}, 
\end{align}
where 
\begin{align}
(M^2)_{11}&=8c_H^2\lambda_1v^2, \\
(M^2)_{22}&=s_H^2(3\lambda_2+\lambda_3)v^2+c_H^2M_1^2-\frac{1}{2}M_2^2,\\
(M^2)_{12}&=\sqrt{\frac{3}{2}}s_H c_H[(2\lambda_4+\lambda_5)v^2-M_1^2]. 
\end{align}
Here $M_1^2$ and $M_2^2$ are introduced to replace $\mu_1$ and $\mu_2$ according to
\begin{align}
M_1^2=-\frac{v}{\sqrt{2}s_H}\mu_1,\quad  M_2^2=-3\sqrt{2}s_H v\mu_2. 
\end{align}
These dimensionful parameters are independent of the VEV, and required in order to take the large mass limit for the 5-plet, 3-plet and singlet ($H_1$) Higgs bosons. 
From the above discussion, the five parameters $\lambda_1$-$\lambda_5$ can be rewritten in terms of physical parameters as: 
\begin{align}
\lambda_1 &=\frac{1}{8v^2c_H^2}(m_h^2c_\alpha^2+m_{H_1}^2s_\alpha^2), \label{lam1} \\
\lambda_2 &=\frac{1}{6v^2s_H^2}
\left[2m_{H_1}^2c_\alpha^2+2m_h^2s_\alpha^2+3M_2^2-2m_{H_5}^2+6c_H^2(m_{H_3}^2-M_1^2)\right],\\
\lambda_3 &= \frac{1}{v^2s_H^2}\left[c_H^2(2M_1^2-3m_{H_3}^2)+m_{H_5}^2-M_2^2\right],  \label{lam3}\\
\lambda_4&= \frac{1}{6v^2s_H c_H}\left[\frac{\sqrt{6}}{2}s_{2\alpha}(m_h^2-m_{H_1}^2)+3s_H c_H(2m_{H_3}^2-M_1^2)\right], \\
\lambda_5&= \frac{2}{v^2}(M_1^2-m_{H_3}^2). \label{lambdas}
\end{align}

The magnitudes of $\lambda$ parameters are theoretically constrained by perturbative unitarity and vacuum stability. 
From Eqs.~(\ref{lam1})-(\ref{lambdas}), these constraints can be translated into bounds on the physical parameters such as the masses of the Higgs bosons and the mixing angles. 
In Refs.~\cite{GM_unitarity,Logan}, the S-wave amplitude matrix has been calculated for all the possible 2-to-2 scatterings of scalar bosons, including the NG bosons and physical Higgs bosons, at high energies. 
By requiring that all the eigenvalues, given as functions of the $\lambda$ parameters, be smaller than a certain value, ({\it e.g.}, 1/2 or 1), we can obtain upper bounds on certain combinations of the $\lambda$ parameters. 
As an independent constraint on the $\lambda$ parameters, the vacuum stability bound is obtained by requiring that the Higgs potential 
be bounded from below in any direction of large scalar boson fields.  In Ref.~\cite{Chiang_Yagyu_GM}, the vacuum stability condition has been derived in all the possible directions with two non-zero scalar fields.

\section{Decays of 5-plet Higgs bosons \label{sec:5plet}}

Since the 5-plet Higgs bosons serve as a distinctive signature of the model, we discuss in this section their decay patterns. 
There are three types of interactions which induce the decays of the 5-plet Higgs bosons at the tree level, namely, 
(i) scalar-gauge-gauge interactions, (ii) scalar-scalar-gauge interactions, and (iii) scalar-scalar-scalar interactions.  
The interaction type (i) is proportional to the triplet VEV $v_\Delta$, and induces the following decay modes:
\begin{align}
H_5^{\pm\pm} \to W^\pm W^\pm,
\quad
H_5^{\pm} \to W^\pm Z,
\quad
H_5^0 \to W^+ W^-~~\text{and}~~ZZ. \label{decay1}
\end{align}
When the mass of the 5-plet Higgs bosons is smaller than the total mass of the final-state gauge bosons in Eq.~(\ref{decay1}), 
one or both of gauge bosons must be off-shell.  In such a case, the above decays should be understood to have three- or four-body final states.  In the following calculations, we consider up to the three-body final states. 
From the interaction type (ii), the following decays are possible as long as kinematically allowed:
\begin{align}
H_5^{\pm\pm} \to W^\pm H_3^\pm,
\quad
H_5^{\pm} \to W^\pm H_3^0~~\text{and}~~ZH_3^\pm,
\quad
H_5^0 \to W^\pm H_3^\mp~~\text{and}~~ZH_3^0. \label{decay2}
\end{align}
The decay rates of these modes are determined by the weak gauge coupling in addition to the masses of the 5- and 3-plet Higgs bosons. 
From the interaction type (iii), the following decays are considered as long as kinematically allowed:
\begin{align}
H_5^{\pm\pm} \to H_3^\pm H_3^\pm,
\quad
H_5^{\pm} \to H_3^\pm H_3^0,
\quad
H_5^0 \to H_3^\pm H_3^\mp~~\text{and}~~H_3^0H_3^0. \label{decay3}
\end{align}
For these decays, the rates depend on the following triple scalar boson couplings:
\begin{align}
\lambda_{H_5^{\pm\pm}H_3^\mp H_3^\mp} &= \frac{v}{\sqrt{2}}\left[(\lambda_3+2\lambda_5)s_H c_H ^2 
+ \frac{\lambda_5}{2} s_H ^3 + \frac{M_1^2}{v^2}s_H ^3  + \frac{M_2^2}{v^2}\frac{c_H ^2}{s_H } \right] \notag\\
& = \frac{i}{\sqrt{2}}\lambda_{H_5^{\pm}H_3^\mp H_3^0} = \sqrt{\frac{3}{2}}\lambda_{H_5^{0}H_3^+ H_3^-}=-\sqrt{\frac{3}{2}}\lambda_{H_5^0 H_3^0 H_3^0},  \label{lam}
\end{align}
where we have defined the above $\lambda_{XYZ}$ couplings as the coefficients of the Lagrangian, ${\cal L}= \lambda_{XYZ}\,XYZ$. 
In addition to the above decay modes, there are also loop induced decays of $H_5^0$, such as $H_5^0 \to \gamma\gamma$ and $H_5^0 \to Z\gamma$. 
Similar to the SM Higgs boson decay, these decays are induced by the $W$ boson loop diagram, but 
have no fermion loop contribution because of the fermiophobic nature of the 5-plet Higgs bosons.  In addition to the $W$ loop contribution, the physical charged Higgs bosons ({\it i.e.}, $H_5^{\pm\pm}$, $H_5^\pm$ and $H_3^\pm$)
also contribute to the decays.  In order to calculate the charged scalar ($H_5^{\pm\pm}$ and $H_5^{\pm}$) loop contributions to the 
$H_5^0 \to \gamma\gamma$ and $H_5^0 \to Z\gamma$ decays, one needs the trilinear couplings:
\begin{align}
\lambda_{H_5^{++}H_5^{--}H_5^0} = -2\sqrt{3}v\left(s_H \lambda_3 -\frac{M_2^2}{3s_H v^2}\right)= -2\lambda_{H_5^{+}H_5^{-}H_5^0}. \label{lam123}
\end{align}
For the $H_3^\pm$ contribution, we use $\lambda_{H_5^0 H_3^+H_3^-}$ given in Eq.~(\ref{lam}).

We note that the $H_5^\pm \to W^\pm Z$ decay is induced by the $H_5^\pm W^\mp Z $ vertex at the tree level. 
One can make a comparison of this vertex with the same vertex in other Higgs-extended models that also has singly-charged Higgs bosons $H^\pm$.  Typically, the magnitude of this vertex is small for the two reasons.  First, if we consider a model whose Higgs sector contains only doublets and singlets, this vertex does not appear at the tree level~\cite{Grifols} but at the loop level. 
As a result, the magnitude of the $H^\pm W^\mp Z$ vertex is not significant.  This vertex has been computed at the one-loop level 
in the two Higgs doublet model (THDM) in Refs.~\cite{hwz1,hwz2,hwz3,ee_HW1}. 
Secondly, if the Higgs sector includes isospin triplets or larger isospin representations, this vertex appears at the tree level.  However, it is proportional to the VEV of such an exotic field, which is strictly constrained by the electroweak rho parameter. 
Therefore, studying this vertex is one good way to identify the GM model. 
The feasibility of examining this vertex has been done for the LHC~\cite{HWZ-LHC} and the ILC~\cite{HWZ-ILC}.

\begin{figure}[t]
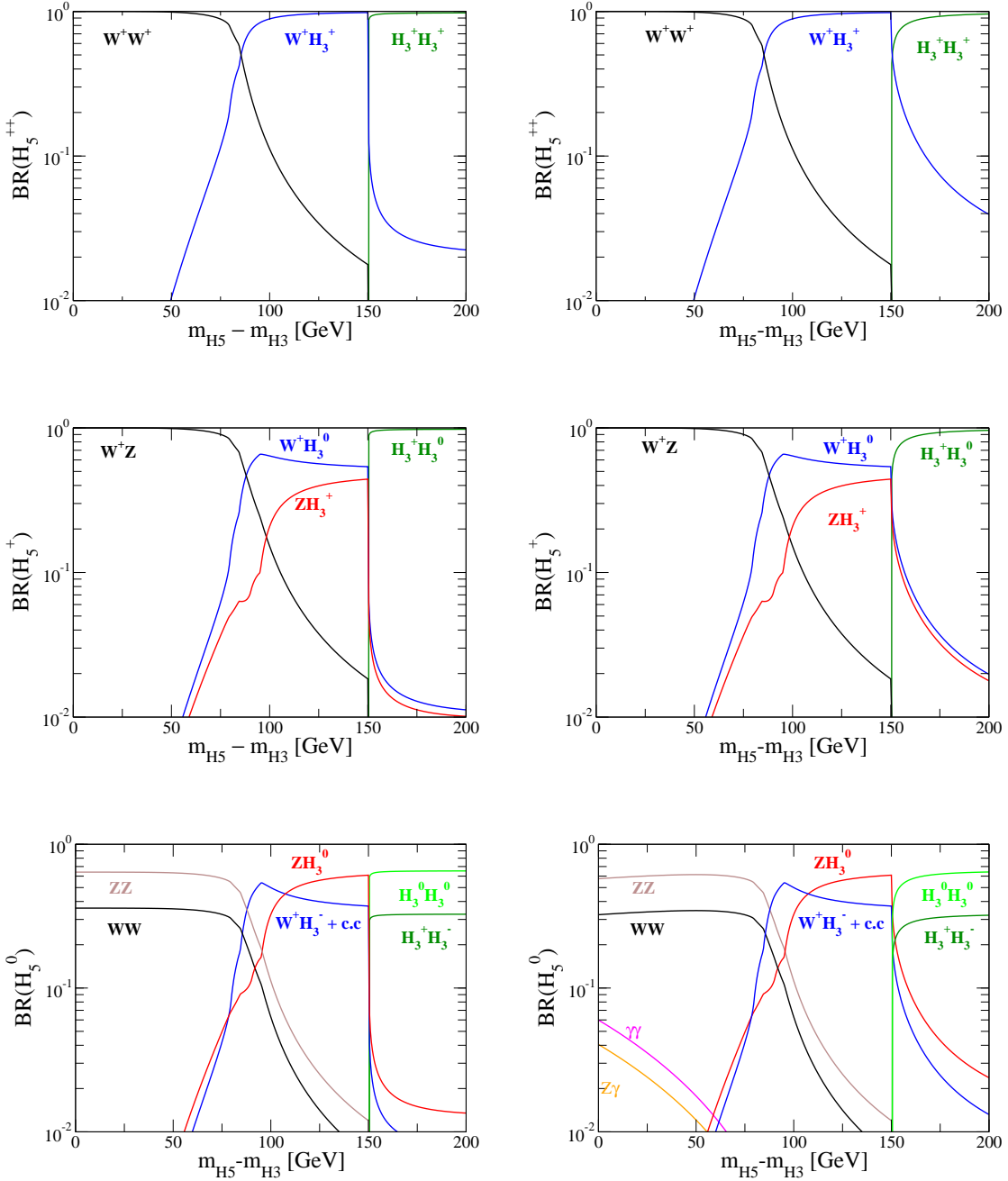

\begin{center}
\includegraphics[width=70mm]{BR_Hpp_300.eps}\hspace{6mm}
\includegraphics[width=70mm]{BR_Hpp_300_2.eps} \\ \vspace{10mm}
\includegraphics[width=70mm]{BR_Hp_300.eps}\hspace{6mm}
\includegraphics[width=70mm]{BR_Hp_300_2.eps} \\ \vspace{10mm}
\includegraphics[width=70mm]{BR_H0_300.eps}\hspace{6mm}
\includegraphics[width=70mm]{BR_H0_300_2.eps}
\caption{
Branching fraction of $H_5^{\pm\pm}$ (upper panels), 
$H_5^{\pm}$ (middle panels) and $H_5^{0}$ (lower panels) as a function of $m_{H_5}^{}-m_{H_3}^{}$ in the case of 
$m_{H_5}^{}=300$ GeV, $v_\Delta = 10$ GeV and $M_2^2=0$. 
The left and right panels show the case with $M_1^2=m_{H_3}^2$ and $M_1^2=0$, respectively.  
}
\label{Fig:decay}
\end{center}
\end{figure}

In Fig.~\ref{Fig:decay}, we show the decay branching ratios of $H_5^{\pm\pm}$, $H_5^{\pm}$ and $H_5^{0}$ as a function of the mass difference $m_{H_5}^{}-m_{H_3}^{}$ in the upper, middle and lower panels, respectively. 
To be specific, we have fixed $m_{H_5}^{}=300$ GeV, $v_\Delta^{} = 10$ GeV and $M_2^2=0$ in all the plots. 
Moreover, $M_1^2=m_{H_3}^2 $ in the left plots and $M_1^2=0$ in the right plots. 
It is seen that the two gauge boson decay modes are dominant when the mass difference is small. 
When the mass difference gets larger, the gauge boson associated decays in Eq.~(\ref{decay2}) and/or
the decays into two scalar bosons Eq.~(\ref{decay3}) become more dominant. 
The difference in the value of $M_1^2$ does not affect the decays of $H_5^{\pm\pm}$ and $H_5^\pm$ much, while 
there appears a small enhancement in the branching ratios for $H_5^0 \to \gamma\gamma$ and $H_5^0 \to Z\gamma$ modes. 
This can be understood by the $\lambda_3$ dependence in the triple scalar boson couplings in Eq.~(\ref{lam123}), where 
a larger value of $\lambda_3$ is given in the case of $M_1^2=0$ as compared to the case of $M_1^2 = m_{H_3}^2$ (see Eq.~(\ref{lam3})). 

\begin{figure}[t]
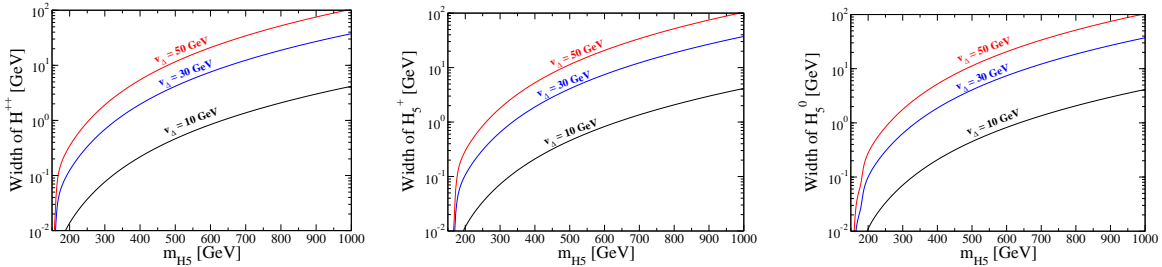

\begin{center}
\includegraphics[width=48mm]{width_Hpp.eps}\hspace{3mm}
\includegraphics[width=48mm]{width_Hp.eps} \hspace{3mm} 
\includegraphics[width=48mm]{width_H0.eps} 
\caption{
Total widths for $H_5^{\pm\pm}$ (left panel)
$H_5^{\pm}$ (center panel), and $H_5^{0}$ (right panel) in the case of $m_{H_3}^{}=M_1=m_{H_5}^{}$ and $M_2=0$. 
The black, blue and red curves respectively show the case of $v_\Delta^{}=10$, 30 and 50 GeV. 
}
\label{Fig:width}
\end{center}
\end{figure}

In Fig.~\ref{Fig:width}, we show the total widths of $H_5^{\pm\pm}$, $H_5^{\pm}$ and $H_5^0$ as functions of $m_{H_5}^{}$. 
In these plots, we take $m_{H_3}^{}=m_{H_5}^{}$, so that only the diboson decays of the 5-plet Higgs bosons are allowed. 
We can see that there is almost no difference among the widths of $H_5^{\pm\pm}$, $H_5^{\pm}$ and $H_5^0$, and that they increase as $m_{H_5}^{}$ gets larger. 
A larger width is also obtained by taking a larger value of $v_{\Delta}^{}$.

We here give a comment on the decay of the 3-plet Higgs bosons. 
An important feature of the 3-plet Higgs bosons is that they have no tree-level scalar-gauge-gauge couplings.  Instead, they have Yukawa couplings that are proportional to $\tan\theta_H$~\cite{Gunion:1989ci} due to the mixing with the Higgs doublet, so that their decay pattern is similar to that of the extra Higgs bosons in the Type-I THDM~\cite{HHG}.  A dedicated study for the production and decays of the 3-plet Higgs boson have been done in Ref.~\cite{Chiang_Yagyu_GM}.

We also would like to mention some characteristic properties of the SM-like Higgs boson $h$ in the model.  First, its couplings with the weak bosons $hVV$ can be larger than those of the SM when 
$v_\Delta^{}\neq 0$ and $\alpha\neq 0$ due to the tree-level mixing with the neutral scalar components from the triplets~\cite{Chiang:2013rua,hvv1,Chiang:2014bia}. 
This feature does not happen in an extended Higgs sector that is only composed of isospin doublets and singlets.  In models with a triplet field or high representations, this is allowed yet with the deviation in the $hVV$ couplings being constrained by the electroweak rho parameter, unless protected by the custodial symmetry as in the GM model. 
Therefore, when the $hVV$ couplings are measured to be larger than the SM predictions in future collider experiments, 
this can be a smoking-gun signature to identify the GM model.

Secondly, sizeable deviations in the decay rates of loop-induced processes, {\it e.g.}, 
$h\to \gamma\gamma$ and $h\to Z\gamma$, are expected due to the $H_5^{\pm\pm}$, $H_5^\pm$ and $H_3^\pm$ loop effects. 
In particular, the correlation between the deviations in 
the decay rates of $h\to \gamma\gamma$ and $h\to Z\gamma$ from the SM values 
gives us a hint for the charged scalar particles running in the loop~\cite{Chiang_Yagyu_GM,Chiang:2012qz} because the two decay modes have different sensitivities to the charged scalars.

\section{Productions of 5-plet Higgs bosons at the ILC \label{sec:production}}

\begin{figure}[t]
\begin{center}
\includegraphics[width=100mm]{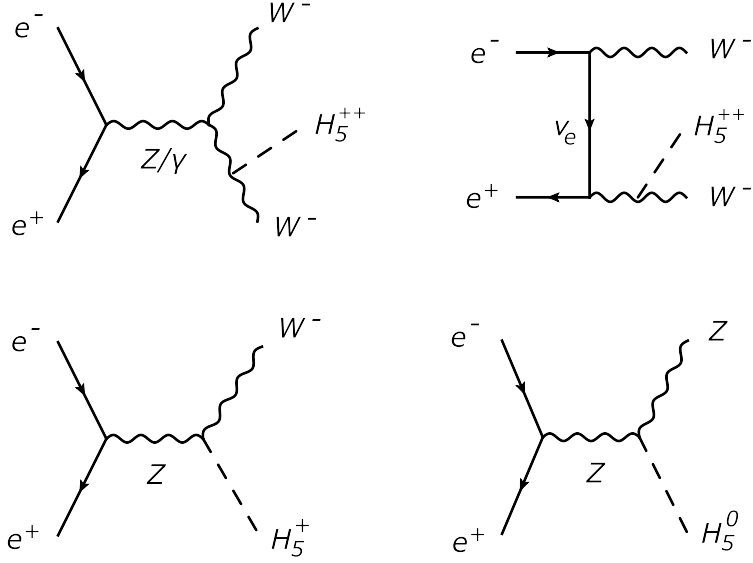}\hspace{6mm}
\end{center}
\caption{
Feynman diagrams of the VBA processes.  }
\label{Fig:VBA}
\end{figure}

\begin{figure}[t]
\begin{center}
\includegraphics[width=100mm]{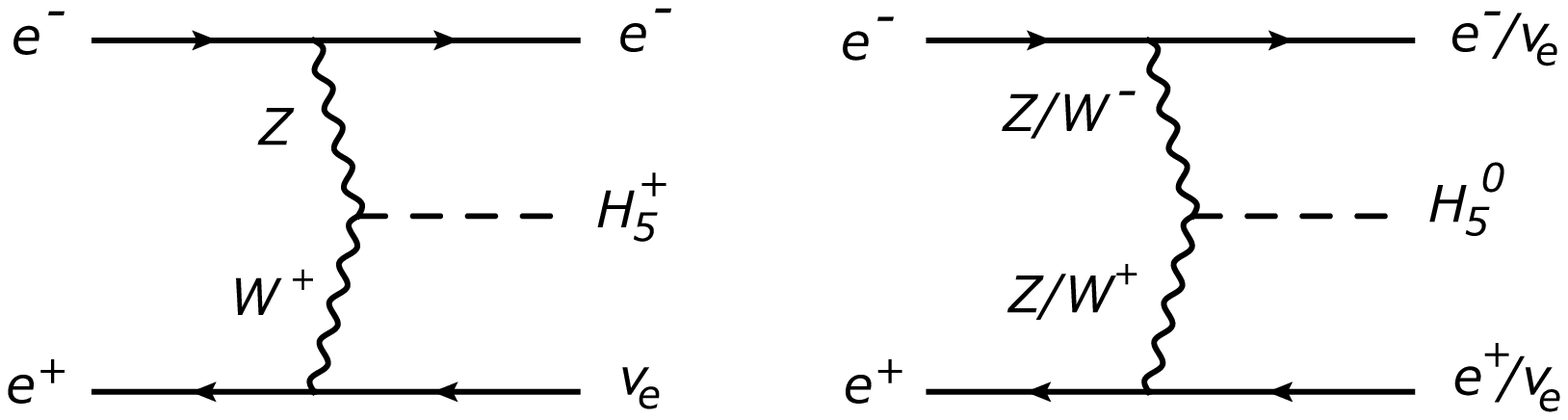}\hspace{6mm}
\end{center}
\caption{
Feynman diagrams of the VBF processes. }
\label{Fig:VBF}
\end{figure}

\begin{figure}[t]
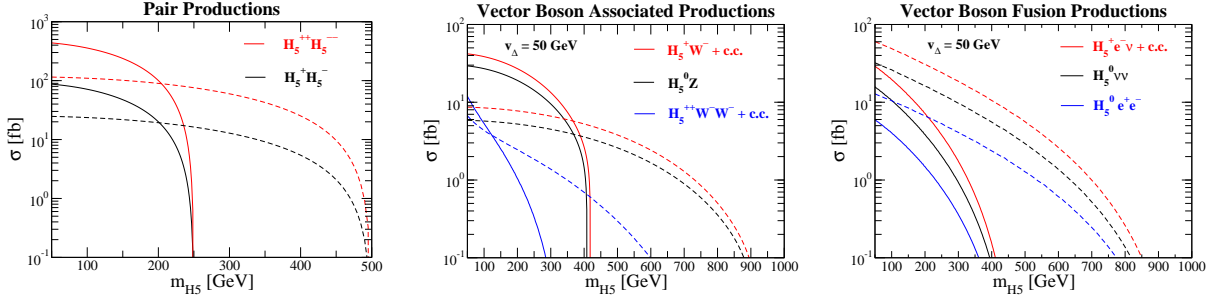

\begin{center}
\includegraphics[width=50mm]{XS_Pair.eps} \hspace{3mm}
\includegraphics[width=50mm]{XS_VBA.eps}\hspace{3mm}
\includegraphics[width=50mm]{XS_VBF.eps}
\caption{
Cross sections of the pair production (left), the VBA (center) and the VBF (right) processes as a function of $m_{H_5}^{}$.  
The collision energy is taken to be 500 GeV (solid curves) and 1 TeV (dashed curves).  For the VBA and VBF processes, we take $v_\Delta=50$ GeV. }
\label{Fig:XS}
\end{center}
\end{figure}

In this section, we discuss the production processes of the 5-plet Higgs bosons at the ILC. 
There are three types of major production modes.  The first type includes pair productions of the doubly-charged and singly-charged Higgs bosons:
\begin{align}
&e^+e^-\to Z^*/\gamma^*\to H_5^{++}H_5^{--}, \\
&e^+e^-\to Z^*/\gamma^*\to H_5^{+}H_5^{-}.
\end{align}
The second type involves the vector boson associated (VBA) processes, as shown in Fig.~\ref{Fig:VBA}:
\begin{align}
&e^+e^- \to H_5^{\pm\pm}W^\mp W^\mp, \\
&e^+e^-\to Z^*\to H_5^{\pm}W^\mp, \\
&e^+e^-\to Z^*\to H_5^{0}Z.
\end{align}
The third type has the vector boson fusion (VBF) processes, as shown in Fig.~\ref{Fig:VBF}:
\begin{align}
&e^+e^-\to  H_5^{\pm}e^\mp \nu_e, \\
&e^+e^-\to  H_5^0 e^+ e^-, \\
&e^+e^-\to  H_5^0 \nu_e \nu_e.
\end{align} 
Among the three type of production modes, 
the cross sections of the VBA and VBF processes depend on $v_\Delta^2$, while that of the pair production is determined solely by the gauge coupling constant. 

It is important to mention here that 
there is a pioneering work by Gunion, Vega and Wudka~\cite{Gunion:1989ci}, in which they calculated the cross sections for pair production, VBF and VBA productions at future $e^+e^-$ colliders. 
In this section, 
we show the production cross sections of these processes
in order to make this paper self-contained 
and to clarify the maximum cross section allowed by the current constraints of the LHC Run-I data. 

In Fig.~\ref{Fig:XS}, the cross sections for the pair production, the VBA and the VBF processes are shown in the left, center and right panels, respectively, at the collision energy $\sqrt{s}=500$~GeV (solid curves) and 1~TeV (dashed curves).  We take $v_\Delta^{}=50$ GeV for the VBA and VBF processes.  The cross sections of these two types of processes for other values of $v_\Delta$ can be obtained readily by scaling with the $v_\Delta$ dependence.

\begin{figure}[t]
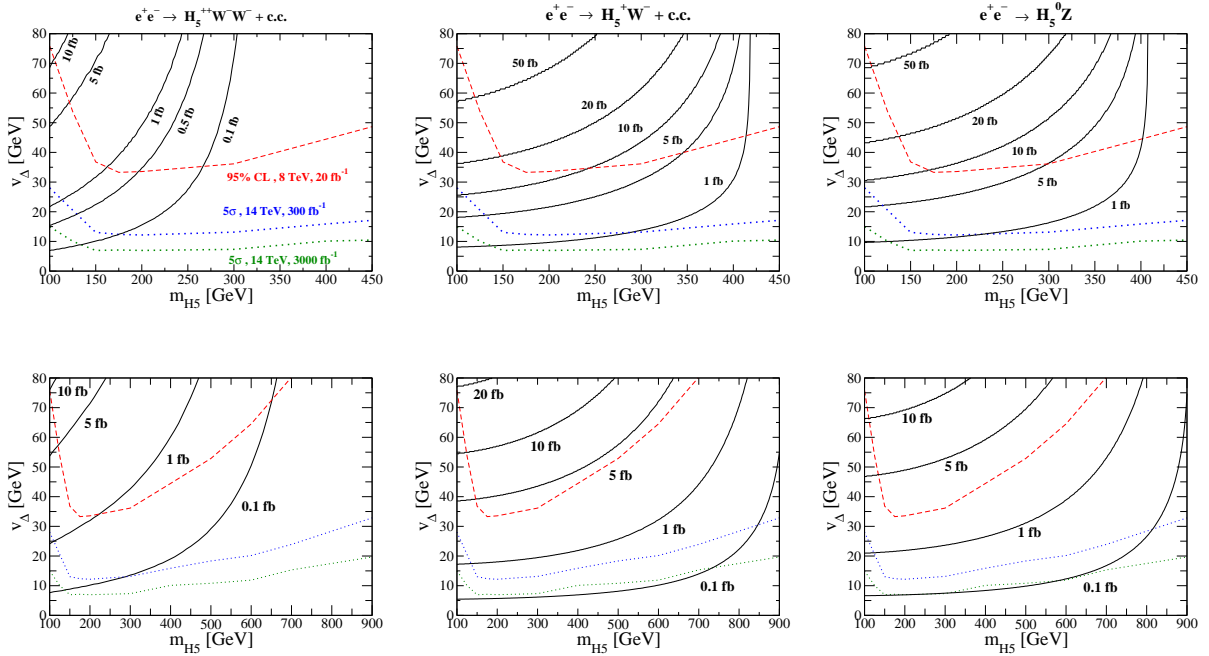

\begin{center}
\includegraphics[width=50mm]{contour_HppWmWm_500.eps}\hspace{3mm}
\includegraphics[width=50mm]{contour_HpWm_500.eps}\hspace{3mm}
\includegraphics[width=50mm]{contour_H0Z_500.eps} \\ \vspace{8mm}
\includegraphics[width=50mm]{contour_HppWmWm_1000.eps}\hspace{3mm}
\includegraphics[width=50mm]{contour_HpWm_1000.eps}\hspace{3mm}
\includegraphics[width=50mm]{contour_H0Z_1000.eps}
\end{center}
\caption{
Contour plots for the VBA production processes.  
The collision energy is taken to be 500 GeV (upper plots) and 1 TeV (lower plots).
The region excluded at the 95\% CL from the LHC Run-I experiment is shown by the red dashed curve, 
and the expected 5-$\sigma$ discovery reach for $H_5^{\pm\pm}$ at the LHC with the collision energy of 14 TeV and 
the integrated luminosity of 300 (3000) fb$^{-1}$ is shown by blue (green) dotted curve. }
\label{Fig:contour1}
\end{figure}

\begin{figure}[t]
\begin{center}
\includegraphics[width=50mm]{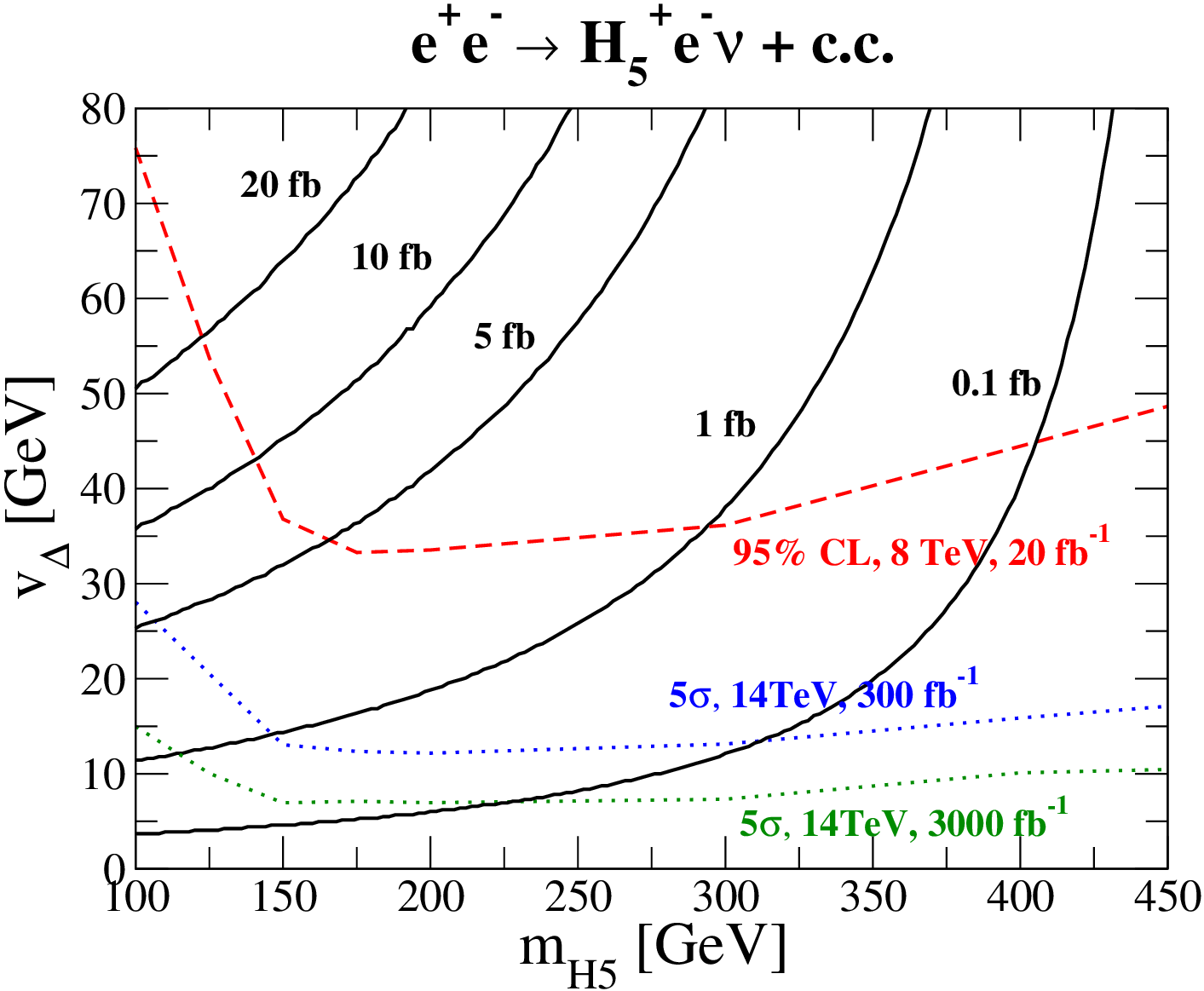}\hspace{3mm}
\includegraphics[width=50mm]{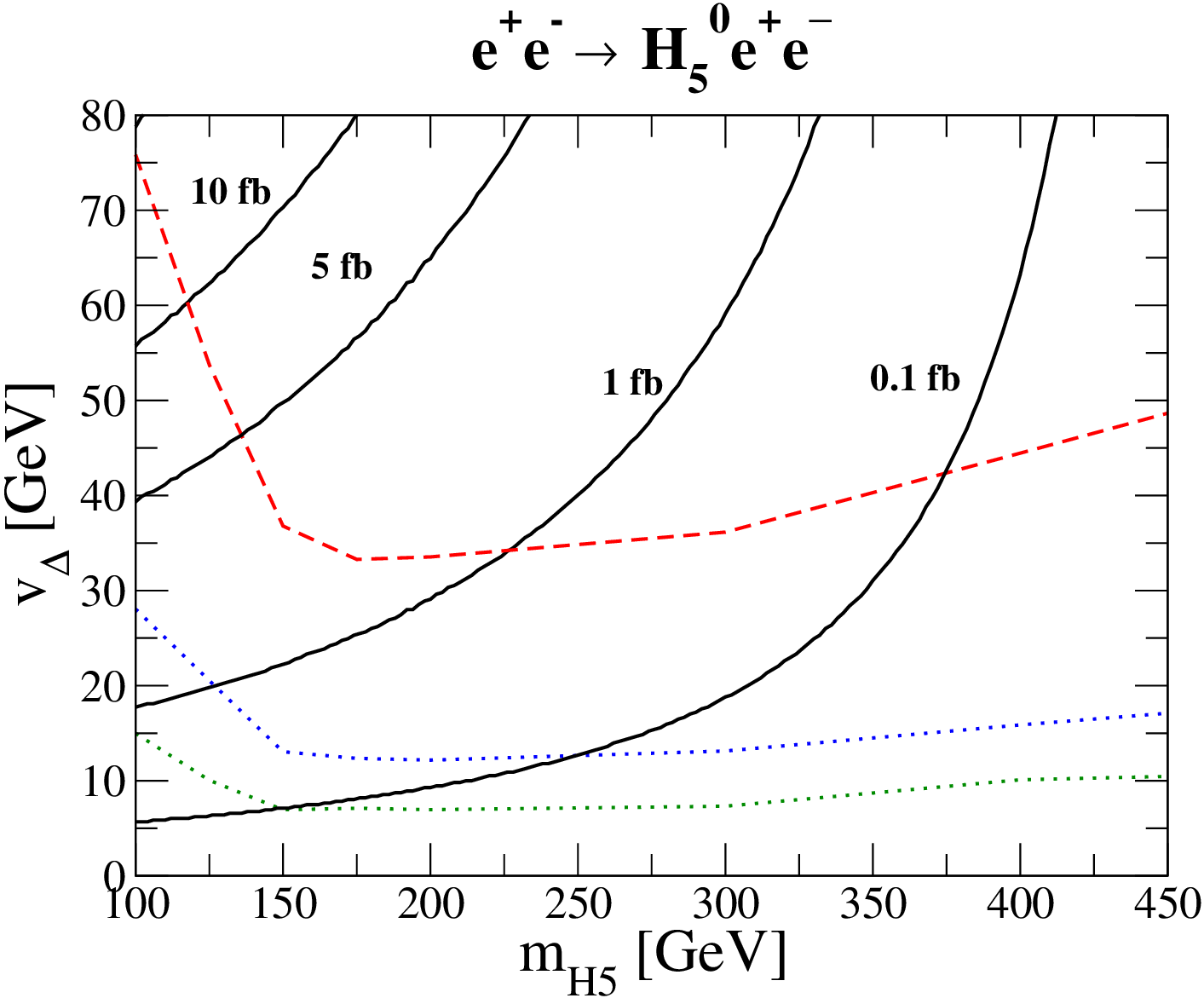}\hspace{3mm}
\includegraphics[width=50mm]{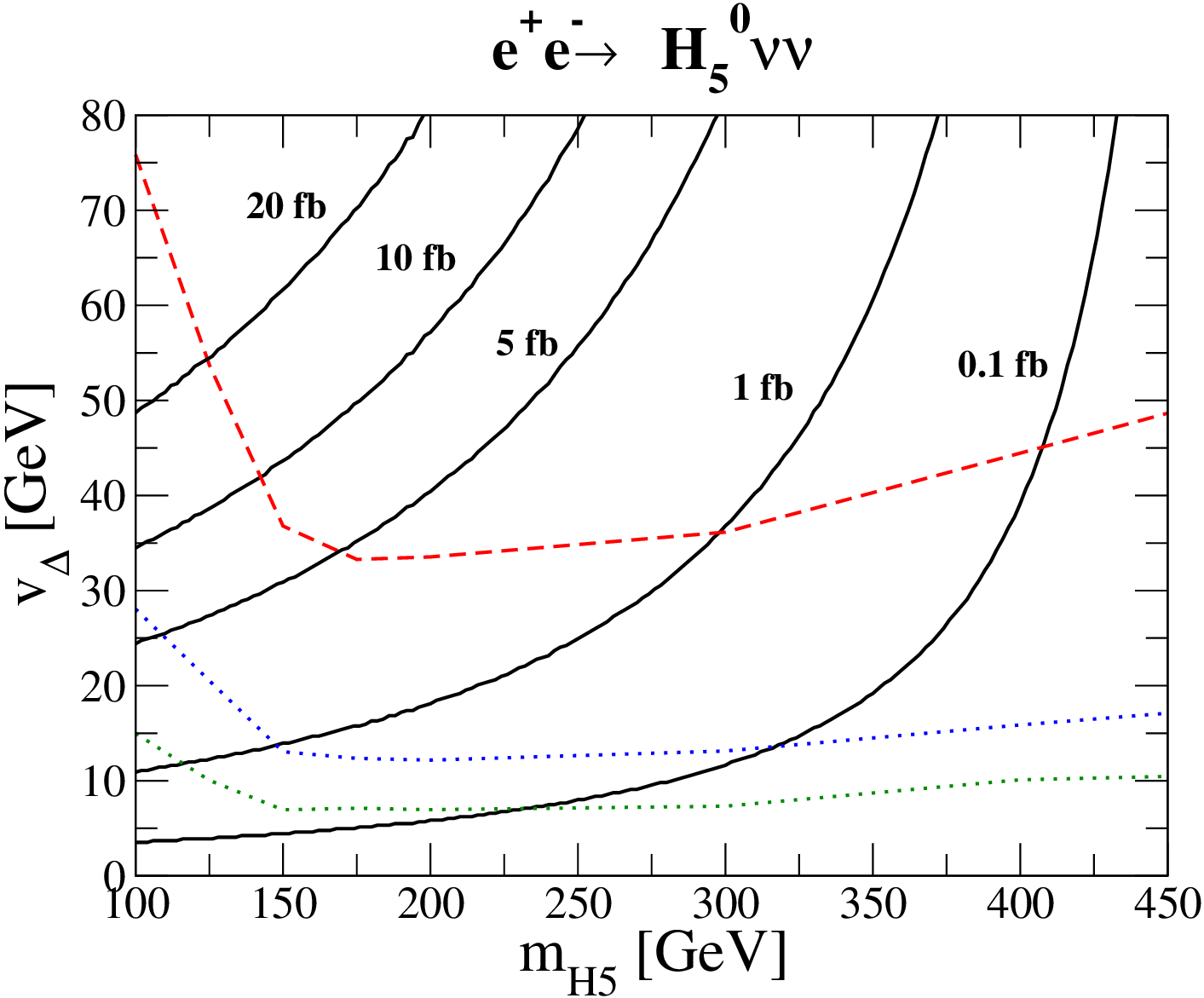}\\ \vspace{8mm}
\includegraphics[width=50mm]{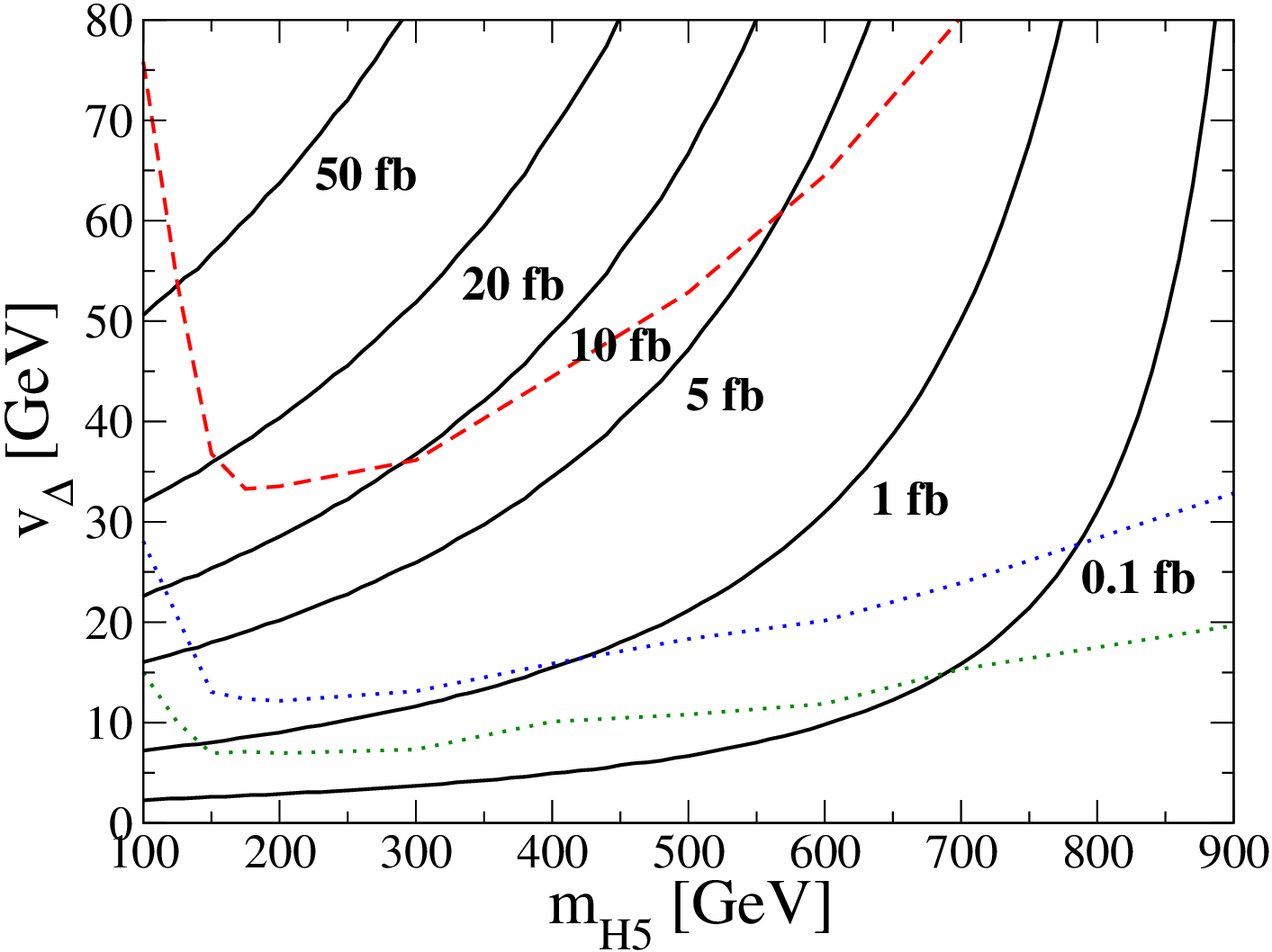}\hspace{3mm}
\includegraphics[width=50mm]{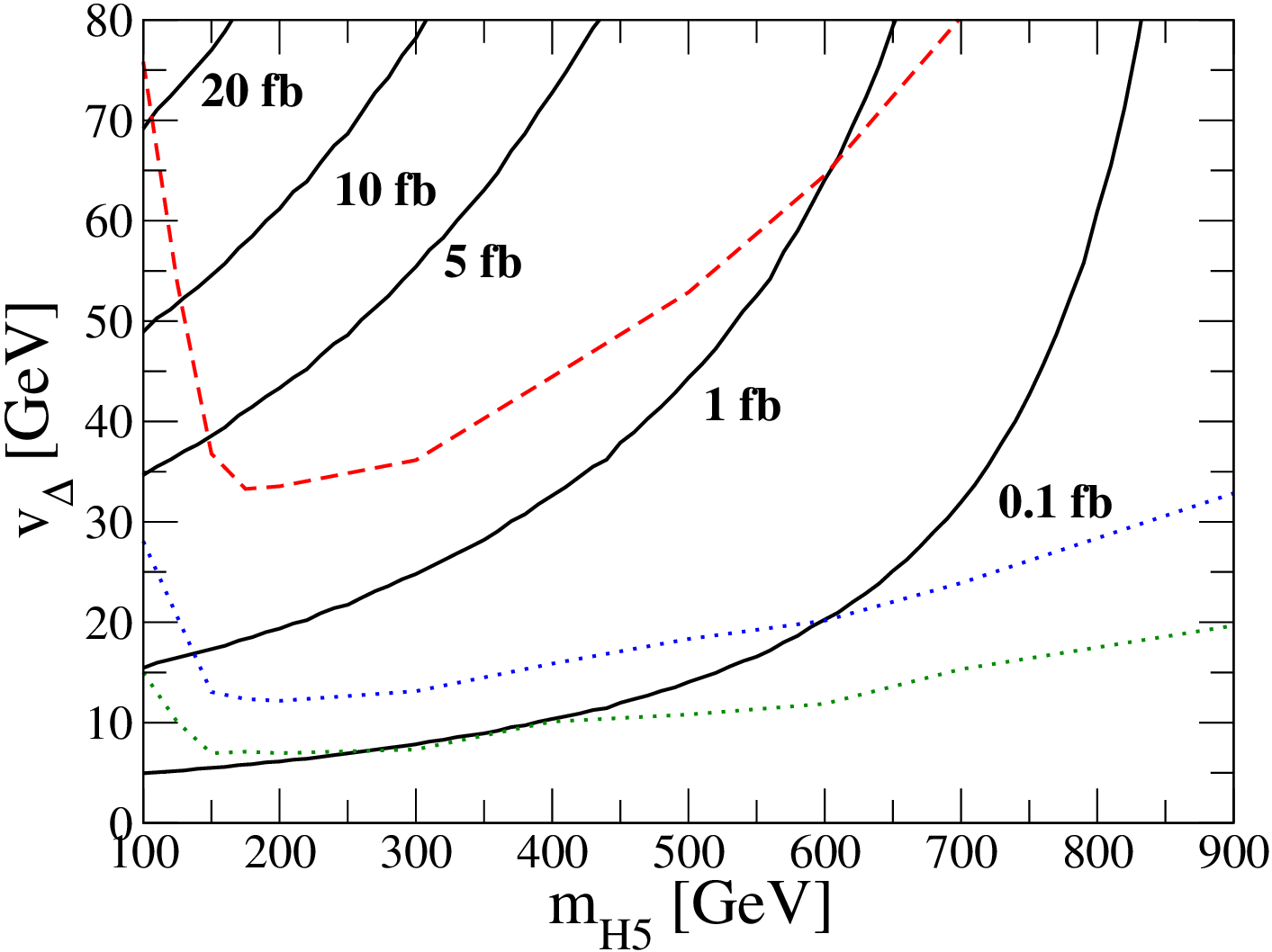}\hspace{3mm}
\includegraphics[width=50mm]{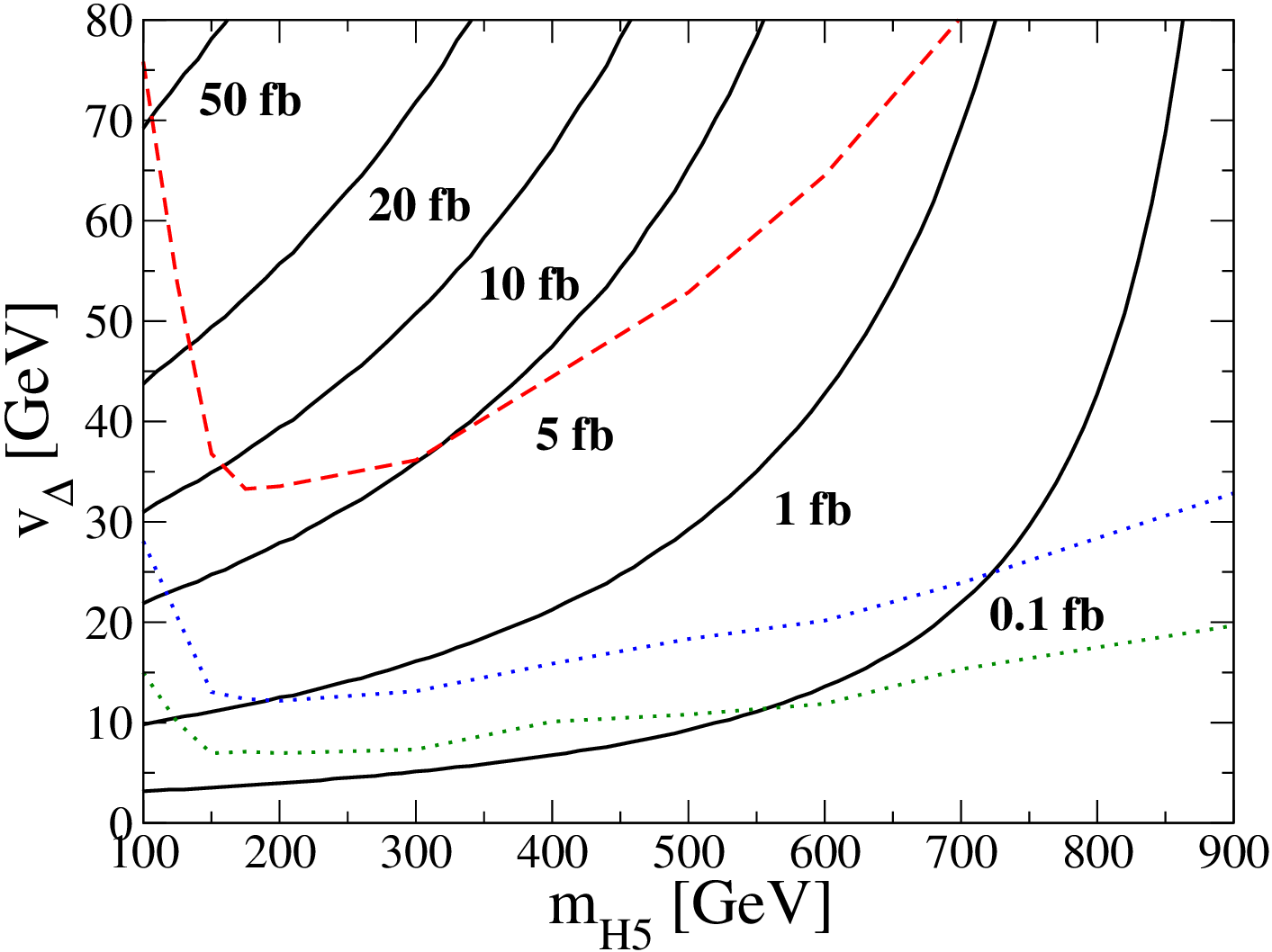}
\end{center}
\caption{
Contour plots for the VBF production processes. 
The collision energy is taken to be 500 GeV (upper plots) and 1 TeV (lower plots).
The region excluded at the 95\% CL from the LHC Run-I experiment is shown by the red dashed curve, 
and the expected 5-$\sigma$ discovery reach for $H_5^{\pm\pm}$ at the LHC with the collision energy of 14 TeV and 
the integrated luminosity of 300 (3000) fb$^{-1}$ is shown by blue (green) dotted curve.
}
\label{Fig:contour2}
\end{figure}

Figs.~\ref{Fig:contour1} and \ref{Fig:contour2} show the contour plots for the cross sections of the VBA and VBF processes, respectively. 
In Ref.~\cite{Chiang:2014bia}, the constraint on the parameter space on the $v_\Delta^{}$-$m_{H_5}^{}$ plane has been studied using the data of same-sign diboson events in the LHC Run-I experiment. 
The current 95\% CL upper bound is indicated in the same plots by red dashed curves.  The $5\sigma$ reach of the 14-TeV LHC with the luminosity of 300~fb$^{-1}$ and 3000~fb$^{-1}$ are drawn in blue and green dotted curves, respectively.

Before closing this section, we comment about productions of the 3-plet Higgs bosons. 
As we discussed in Sec.~III, the 3-plet Higgs bosons have the fermion-specific nature; that is, 
there are $H_3 f\bar{f}$ couplings but no $H_3 VV$ couplings. 
Therefore, their possible production mechanisms at the ILC are the pair production $e^+e^- \to H_3^+ H_3^-$ and the fermion associated processes $e^+e^- \to f\bar{f}H_3^0$ and $e^+e^- \to f\bar{f}'H_3^\pm$. 
In Ref.~\cite{Zheng}, a comprehensive analysis on the production and decay processes of the extra Higgs bosons for four types of Yukawa interactions in the THDM under a softly-broken $Z_2$ symmetry has been performed.  Due to the similarity between the 3-plet Higgs bosons and the extra Higgs bosons in the Type-I THDM, production cross sections similar to those for the Type-I THDM in Ref.~\cite{Zheng} are expected for the 3-plet Higgs bosons.

\begin{center}
\begin{table}[t]
\begin{tabular}{c|c|c|c|c|c}\hline \hline
$\sqrt{s}$&$ZZZ$ & $W^+W^-Z$ & $W^+W^-W^+W^-$ & $W^+W^-ZZ$ & $ZZZZ$ \\ \hline  
500 GeV&1.1 fb & 39 fb & 0.13 fb & 0.036 fb & $6.8\times 10^{-4}$ fb \\ \hline  
1 TeV& 0.86 fb & 57 fb & 0.79 fb & 0.46 fb & $3.0\times 10^{-3}$ fb \\ \hline  \hline
\end{tabular} 
\caption{Cross sections for the 3- and 4-gauge boson final states in the SM with $\sqrt{s}=500$~GeV and 1~TeV.  
}
\label{multi-gauge}
\end{table}
\end{center}

In Table~\ref{multi-gauge}, we show the cross sections of 3- and 4-gauge boson final states in the SM. 
These cross sections are calculated by using {\tt CalcHEP\_3.4.2}~\cite{calchep}. 
Among them, the $W^+W^-Z$ channel is the most relevant to the analysis given in the next section.  Its cross section is typically one order of magnitude larger than the maximally allowed value of the signal cross section.

\section{Reconstruction of 5-plet Higgs bosons from $W^+W^-Z$ events \label{sec:reconstruction}}

We now show various distributions for the $e^+e^-\to W^+W^-Z$ process, where 
there are two 5-plet Higgs contributions: $e^+e^-\to H_5^0 Z \to W^+W^-Z$ and $e^+e^-\to H_5^\pm W^\mp \to W^+W^-Z$. 
In this section, we assume $m_{H_3}>m_{H_5}$, where 
the branching fraction of the $H_5 \to VV$ decay modes becomes 100\%, because the $H_5 \to V^{(*)}H_3$ and $H_5 \to H_3 H_3$ modes are kinematically forbidden. 
The production cross sections for these processes are given by 
\begin{align}
\sigma(e^+e^- \to Z^* \to H_5^0 Z) 
&= \frac{g_Z^6v_\Delta^2}{32\pi s^2}\frac{8}{3}\frac{v_e^2 + a_e^2}{(1-x_Z^{})^2}\lambda^{1/2}(x_Z^{},x_{H_5})\notag\\
& \qquad \times \left\{ 1 + \frac{1}{4x_Z^{}}[(1+x_Z^{}-x_{H_5})^2-\frac{1}{3}\lambda(x_Z^{},x_{H_5})]\right\}, \\
\sigma(e^+e^- \to Z^* \to H_5^\pm W^\mp) 
&= \frac{2g_Z^4g^2v_\Delta^2}{32\pi s^2}\frac{v_e^2 + a_e^2}{(1-x_Z^{})^2}\lambda^{1/2}(x_W^{},x_{H_5})\notag\\
& \qquad \times \left\{ 1 + \frac{1}{4x_W^{}}[(1+x_W^{}-x_{H_5})^2-\frac{1}{3}\lambda(x_W^{},x_{H_5})]\right\}, 
\end{align}
where $x_A^{} = m_A^2/s$, $v_e = -1/4+s_W^2$, and $a_e =-1/4$. 
The phase-space function $\lambda$ is given by $\lambda(x,y)=1+x^2+y^2-2x-2y-2xy$.

Consider 2-to-2 scattering processes $e^+e^-\to P_1 P_2$, where $P_1$ and $P_2$ denote particles with masses of $m_1$ and $m_2$ and energies of $E_1$ and $E_2$, respectively.  The energies are explicitly given by 
\begin{align}
E_1^{} = \frac{\sqrt{s}}{2}(1+x_1^{}-x_2),\quad
E_2^{} = \frac{\sqrt{s}}{2}(1+x_2^{}-x_1), \quad (E_1+E_2=\sqrt{s}),  \label{recoil}
\end{align}
where $x_i=m_i^2/s$.  As a benefit of the ILC, we have the information of the initial collision energy $\sqrt{s}$.  Therefore, by measuring $E_1$ and $E_2$, one can reconstruct the masses $m_1$ and $m_2$ using Eq.~(\ref{recoil}) without ambiguity.

\begin{figure}[!t]
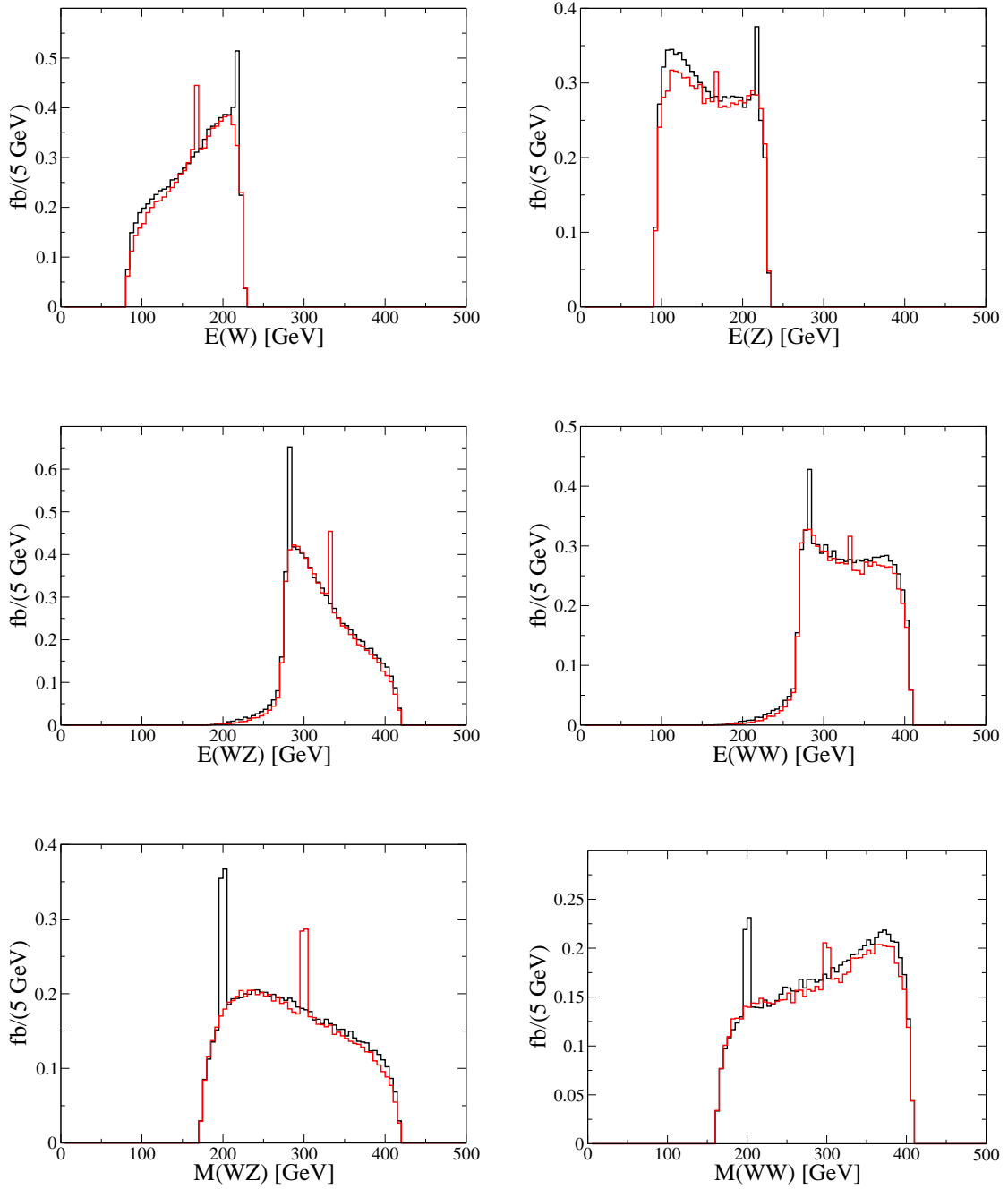

\begin{center}
\includegraphics[width=70mm]{EW_500.eps}\hspace{6mm} 
\includegraphics[width=70mm]{EZ_500.eps}\\
\vspace{10mm}
\includegraphics[width=70mm]{EWZ_500.eps}\hspace{6mm}
\includegraphics[width=70mm]{EWW_500.eps}\\
\vspace{10mm}
\includegraphics[width=70mm]{MWZ_500.eps}\hspace{6mm} 
\includegraphics[width=70mm]{MWW_500.eps}
\end{center}
\caption{
Energy (upper and center panels) and invariant mass (lower panels) 
distributions for the $e^+e^-\to W^+W^-Z$ process in the case of $\sqrt{s}=500$~GeV and $v_\Delta = 30$~GeV, including the ISR with the energy scale at $\sqrt{s}$.
We take $m_{H_5}=200$ GeV (black) and 300 GeV (red). 
In all of these plots, both SM background processes and signal process are included. 
}
\label{Fig:distr500}
\end{figure}

\begin{figure}[!t]
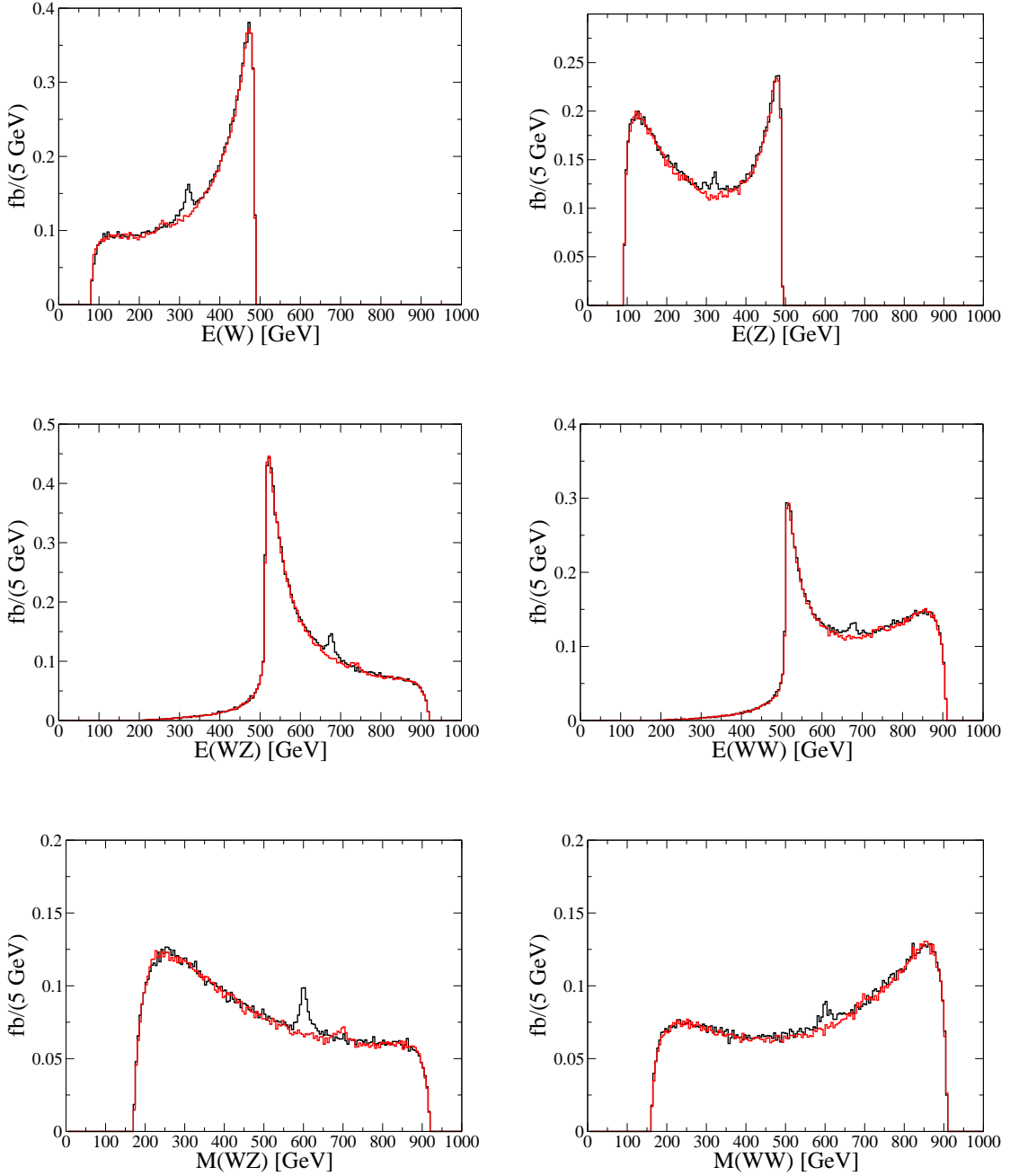

\begin{center}
\includegraphics[width=70mm]{EW_1000.eps}\hspace{6mm} 
\includegraphics[width=70mm]{EZ_1000.eps}\\
\vspace{10mm}
\includegraphics[width=70mm]{EWZ_1000.eps}\hspace{6mm}
\includegraphics[width=70mm]{EWW_1000.eps}\\
\vspace{10mm}
\includegraphics[width=70mm]{MWZ_1000.eps}\hspace{6mm} 
\includegraphics[width=70mm]{MWW_1000.eps}
\end{center}
\caption{
Same as Fig.~\ref{Fig:distr500}, but for $\sqrt{s} = 1$~TeV and $v_\Delta^{}=50$~GeV.  Here we take $m_{H_5} = 600$~GeV (black) and 700~GeV (red).  }
\label{Fig:distr1000}
\end{figure}

In Fig.~\ref{Fig:distr500}, we show various distribution plots for the $e^+e^-\to W^+W^-Z$ process.  We take $m_{H_5}^{}=200$ (black curve) and 300~GeV (red curve) and the collision energy $\sqrt{s}=500$ GeV and $v_\Delta = 30$~GeV.  The same are plot in Fig.~\ref{Fig:distr1000}, but for $\sqrt{s} = 1$~TeV, $v_\Delta^{}=50$~GeV and $m_{H_5} = 600$~GeV (black curve) and 700~GeV (red curve).
We use {\tt CalcHEP\_3.4.2}~\cite{calchep} for this analysis, where the effect of initial state radiation (ISR) is 
taken into account with the fixed ISR energy scale at $\sqrt{s}$.  In each of the figures, the upper left plot shows the distribution in the energy of $W$ boson, $E_W^{}$; the upper right plot shows the distribution in the energy of $Z$ boson, $E_Z^{}$; the center left plot shows the distribution in the total energy of the $WZ$ system, $E_{WZ}^{}$; the center right plot shows the distribution in the total energy of the $WW$ system, $E_{WW}^{}$; the lower left plot shows the distribution in the invariant mass of the $WZ$ system, $M_{WZ}^{}$; and the lower right plot shows the distribution in the invariant mass of the $WW$ system, $M_{WW}^{}$. 
We note that the distribution for $W$ $(WZ)$ is for either $W^+$ or $W^-$ ($W^+Z$ or $W^-Z$), as there is no difference between the two.  
In both $M_{WZ}$ and $M_{WW}$ distributions, 
the peak appears at around $m_{H_5}^{}$ due to the contribution from $H_5^\pm$ and $H_5^0$, respectively. 
In the case of $m_{H_5}^{}=700$ GeV, it is difficult to find a peak (see the bottom panels in Fig.~\ref{Fig:distr1000}). 
The reason is that the signal cross section is suppressed, while the widths of $H_5^\pm$ and $H_5^0$ become large 
for a larger value of $m_{H_5}^{}$. 
Thus, the height and width of the peak in the invariant mass distribution becomes smaller and broader, respectively. 
From the simultaneous observation of the two peaks in $M_{WZ}$ and $M_{WW}$ at the same position, 
we can test the degeneracy of $H_5^\pm$ and $H_5^0$ in mass, which serves as the evidence for the custodial symmetry in the GM model.

\section{Discussions \label{sec:discussions}}

We now discuss how the searches for the 5-plet Higgs bosons at the ILC, as investigated in this paper, can be 
complementary to those at the LHC.

At the LHC, the most promising process for $H_5^{\pm\pm}$ is the VBF processes, {\it i.e.}, $qq \to q'q'W^\pm W^\pm \to q'q' H_5^{\pm\pm}\to q'q' W^\pm W^\pm$. 
In order to obtain a sufficiently large cross section to discover $H_5^{\pm\pm}$ in this process, 
we need a large triplet VEV, because the $H_5^{\pm\pm}W^\mp W^\mp$ vertices are proportional to $v_\Delta^{}$. 
In the GM model, such a large value is allowed without conflict with the experimental data as alluded to in Sec.~II. 
The parameter region which allows the 5-$\sigma$ discovery of $H_5^{\pm\pm}$ is shown in Fig.~\ref{Fig:contour1} by the blue (green) dotted curve, where the collision energy and the integrated luminosity are taken to be 14~TeV and 300 (3000)~fb$^{-1}$, respectively. 
To obtain these discovery reaches, leptonic decays of the same-sign $W$ boson are assumed. 
For example, a 5-$\sigma$ discovery is expected by taking $v_\Delta^{} \gtrsim 17~(20)$~GeV and $m_{H_5}^{}=500~(800)$~GeV, assuming the collision energy of 14 TeV and the integrated luminosity of 300 fb$^{-1}$.

For the detection of $H_5^\pm$ and $H_5^0$, one can use the similar VBF processes such as 
$qq \to q'q'W^\pm Z \to q'q' H_5^{\pm}\to q'q' W^\pm Z$ and 
$qq \to q'q'W^+ W^-/ZZ \to q'q' H_5^{0}\to q'q' W^+ W^-/ZZ$. 
It must be emphasized here that the above-mentioned processes are not significant in the minimal Higgs triplet model because of the strong restriction on the triplet VEV from the electroweak rho parameter. 
The discovery reaches for these processes are worse than that of $H^{\pm\pm}_5$ because of larger cross sections in the SM backgrounds. 
A dedicated simulation study for these VBF processes had been done in Ref.~\cite{Chiang_Yagyu_GM}, and it was shown that the significances of the signatures via the $H_5^\pm$ and $H_5^0$ productions are smaller than that via the $H_5^{\pm\pm}$ production. 
For example, in the case of $m_{H_5}=140$ GeV and $v_\Delta^{}=20$ GeV, the signal significance has been given to be about $23$, $8.2$ and $3.9$ for the collision energy of 14 TeV and 
the integrated luminosity of 100~fb$^{-1}$ after imposing appropriate kinematic cuts~\cite{Chiang_Yagyu_GM}. 
In the analysis, the leptonic decays of $W$ and $Z$ bosons were assumed. 
As noted in Sec.~III, the SM-like Higgs boson coupling $hVV$ can be larger than the corresponding SM value when $v_\Delta\neq 0$ and $\alpha\neq 0$. 
If such an enhancement is realized, the cross section of the VBF process mediated by the neutral Higgs bosons ({\it i.e.}, $h$, $H_1^0$ and $H_5^{0}$) would also be enhanced, rendering a larger signal significance for the process. 
The VBF processes with an enhanced $hVV$ coupling had been discussed in Ref.~\cite{Chiang:2013rua}, and were found to be promising for discovering the 5-plet Higgs bosons at the LHC.

A further test for the identification of the GM model is to check the mass degeneracy among $H_{5}^{\pm\pm}$, $H_5^\pm$ and $H_5^0$.  In order to reconstruct the masses of these Higgs bosons, it is better to use the hadronic decays of the weak bosons from the decays of the 5-plet Higgs bosons.  
The energy resolution for the dijet system turns out to be important for the reconstruction. 
In particular, the ability to discriminate the dijet event from a $W$ boson and a $Z$ boson is crucial in the test of mass degeneracy. 

Now, let us discuss the value of ILC experiments for testing the GM model after the LHC experiments. 
One of the most important advantages at the ILC is the good energy resolution for jet systems. 
At the ILC, the target energy resolution $\sigma_E^{}$ for a dijet system is 
$\sigma_E^{} = 0.3\times \sqrt{E_{jj}}$~GeV~\cite{ILC-TDR4} where $E_{jj}$ is the dijet energy.  Therefore, $\sigma_E^{}\simeq 3$ GeV for $E_{jj}\simeq 100$ GeV, which
allows us to distinguish dijets from $W$ and $Z$. 
Because of this detector performance, 
there are the following two advantages to test the GM model at the ILC. 
First,  
the precise measurement of the $H_5^\pm W^\mp Z$ vertex (or more generally the $H^\pm W^\mp Z$ vertex for physical singly-charged Higgs bosons $H^\pm$)
is possible via the $e^+e^- \to Z^* \to H^\pm W^\mp$ process. 
In Ref.~\cite{HWZ-ILC}, the feasibility of measuring the $H^\pm W^\mp Z$ vertex has been discussed by using the recoil method; {\it i.e.}, 
the reconstruction of a hadronic $W$ boson decay, where only leptonic decays
of $H^\pm$ were considered. 
Second, the good dijet energy resolution makes 
the analysis of the $e^+e^- \to W^+W^-Z$ process discussed in Sec.~V realistic. 
As discussed in Sec.~V, 
the observation of the distinctive peaks at the same position in the invariant mass distributions of $M_{WW}$ and $M_{WZ}$
indicates that there are neutral and singly-charged particles with degenerate mass.  
Testing the mass degeneracy can be the direct evidence for these particles to be identified as the 5-plet Higgs bosons $H_5^\pm$ and $H_5^0$ for the GM model. 

Finally, we would like to comment on 
a signal and background simulation of the $e^+e^-\to W^+W^-Z$ process 
with the actual final state such as multi-lepton plus jets at a detector level, which is not performed in this paper. 
Such an analysis is needed to clarify the feasibility of the method proposed in this paper to test the masses of 5-plet Higgs bosons, and 
that would be best done by the experimental colleagues who have full information about detector design and efficiency.

\section{Conclusions \label{sec:summary}}

We studied in this work the phenomenology of exotic Higgs bosons in the Georgi-Machacek (GM) model in the environment of the ILC, 
assuming the colliding energies of 500~GeV and 1~TeV.  
We showed how the decay branching ratios of the three charged states of 5-plet Higgs fields depend on the mass difference between the 5-plet and the 3-plet.  
It was found that except in the large mass splitting regime, the branching ratios did not change much as the parameter $M_1^2$ varies from $m_{H_3}$ to $0$, 
except that the $H_5^0 \to \gamma\gamma/Z\gamma$ decay rates became larger in the latter case.  
It was also noted that the $h \to \gamma\gamma/Z\gamma$ decay rates were expected to differ from their SM values due to the participation of the charged Higgs bosons in the loop.  

We then studied the production of the 5-plet Higgs bosons at the ILC, {\it i.e.}, 
the pair production, the vector boson associated production and the vector boson fusion production at $\sqrt{s} = 500$~GeV and 1 TeV. 
While the pair productions can only be used in the case of $m_{H_5}^{}<\sqrt{s}/2$,  
the vector boson associated and vector boson fusion processes are valid even when the mass is larger than $\sqrt{s}/2$. 
In particular, we found that the vector boson associated production had a bigger cross section than the vector boson fusion processes at $\sqrt{s}=500$ GeV.
For $\sqrt{s}=1$~TeV, the cross section of the vector boson fusion process is larger than that of the vector boson associated process, especially in the small mass region, {\it e.g.}, $m_{H_5}^{}\lesssim 500$~GeV, while they become comparable when $m_{H_5}^{}\gtrsim 500$~GeV. 

We showed explicitly that with a cleaner collider environment, 
it was easier to determine from various energy and invariant mass distributions of the $W^+ W^- Z$ final state the masses of $H_5^{\pm}$ and $H_5^0$ 
with high precision at the ILC than the LHC.  
Combined with the information of the $H_5^{\pm\pm}$ mass, 
one would be able to have a comprehensive test of the mass degeneracy within the 5-plet, thereby identifying the GM model.  
Finally, we discussed how the ILC study of the Higgs bosons in the GM model would complement that at the LHC.

\section*{Acknowledgments}

This work was supported in part by the Ministry of Science and Technology, Taiwanunder Grant Nos. MOST-100-2628-M-008-003-MY4 and 104-2628-M-008-004-MY4.  
SK's work was supported, in part, by Grant-In-Aid for Scientific Research, No. 23104006,
and by Grant H2020-MSCA-RISE-2014, No. 645722 (Multi-Higgs).
KY was supported by JSPS postdoctoral fellowships for research abroad.

\end{document}